\pdfoutput=1
\documentclass[pra,10pt,twocolumn,amsmath,amssymb,nofootinbib,superscriptaddress]{revtex4}
\usepackage[english]{babel}  
\usepackage{mathrsfs,amsfonts,graphics}

\def\bra#1{\mathinner{\langle{#1}|}}
\def\ket#1{\mathinner{|{#1}\rangle}}
\def\braket#1{\mathinner{\langle{#1}\rangle}}
\def\Bra#1{\left\langle#1\right|}
\def\Ket#1{\left|#1\right \rangle}
{\catcode`\|=\active 
  \gdef\Braket#1{\begingroup
\mathcode`\|32768\let|\BraVert\left<{#1}\right>\endgroup}}
\def\BraVert{\egroup\,\mid\,\bgroup}

\usepackage{cancel}
\usepackage{color}
\definecolor{Blue}{rgb}{0,0,1}
\definecolor{Red}{rgb}{1,0,0}
\definecolor{Green}{rgb}{0,1,0}
\definecolor{Purp}{rgb}{.2,0,.2}
\definecolor{white}{rgb}{1,1,1}

\begin{document}

\title{The role of preparation in quantum process tomography}

\author{Kavan Modi}
 \email{modikk@gmail.com}
\affiliation{Centre for Quantum Technologies, National University of Singapore, Singapore}
\affiliation{Center for Complex Quantum Systems, Department of Physics, The University of Texas at Austin, Austin TX, USA}
\author{E.~C.~G. Sudarshan}
\affiliation{Center for Complex Quantum Systems, Department of Physics, The University of Texas at Austin, Austin TX, USA}

\begin{abstract}
In a recent letter one of us pointed out how differences in preparation procedures for quantum experiments can lead to non-trivial differences in the results of the experiment.  The difference arise from the initial correlations between the system and environment.  Therefore, any quantum experiment that is prone to the influences from the environment must be prepared carefully.  In this paper, we study quantum process tomography in light of this.  We suggest several experimental setups, where preparation of initial state plays a role on the final outcome of the experiment.  We show that by studying the linearity and the positivity of the resulting maps the experimenter can  determine the nature of the initial correlations between the system and the environment.
\end{abstract}

\date{\today}
\maketitle

Quantum information processing promises powerful computational methods that surpass the methods of classical information processing \cite{Nielsen00a,  shannon48a}. These methods rely on taking advantage of quantum parallelism by using quantum superposition and quantum entanglement as resources.  In order to implement such a device one must have precise control over the system, and isolate it from the surrounding environment to preserve coherence.  Yet realistically, it is nearly impossible to isolate the system of interest completely from its surroundings, while having a great deal of control.

With the rising interest in quantum computation and quantum information processing, quantum coherence experiments are performed readily these days, though with relatively small systems.  One of the major problem with these experiment is the loss of coherence due to the interaction between the system of interest and the unknown environmental states. The methods for studying the interaction between the system and the environment are given by the quantum theory of open systems. 

The quantum theory of open systems got its start in almost fifty years ago with the introduction of dynamical maps \cite{SudarshanMatthewsRau61, SudarshanJordan61} due to Sudarshan, Mathews, Rau, and Jordan.  Decades after its conception, the dynamical map formalism is finally being tested in the laboratory setting.  The experimental determination of a dynamical map is achieved by a procedure called \emph{quantum process tomography}.

Any experiment, including quantum process tomography experiments, requires a method to prepare the initial states of the system at the beginning of the experiment \cite{kuah:042113,kuahdis,modidis,preppap}.  We study the affects of the preparation procedure on quantum systems that interact with an environment.  The act of preparation has been neglected from the theory of quantum process tomography (and for all quantum experiments that interact with a non-trivial environment).  We investigate this issue for quantum process tomography in detail in this paper. We present several simple examples to motivate our arguments.  

In previous papers we studied not completely positivity of dynamical maps as function of initial correlations between the system and the environment \cite{CesarEtal07} and the role of state preparation in quantum process tomography \cite{kuah:042113} (and quantum mechanics in \cite{preppap}).  This paper is an extension of those studies. 

In Sec. \ref{secqpt}, we review the concepts and the mathematics of quantum process tomography and preparation procedure.  In Sec. \ref{stochprep}, we construct and discuss several simple examples of quantum process tomography for different preparations procedures.   In Sec. \ref{secexp}, we follow these examples up with analyzing some recent quantum process tomography experiment.  The example are in the same spirit as the ones in the previous papers, therefore allowing us to compare all of the results.  We reproduce the examples from previous papers in the Appendix of this paper for ease.  We have modified the language of these examples to fit the language of this paper.
And finally in Sec. \ref{seccon}, we discuss how preparation procedures differentiate between the outcome of a quantum process tomography experiment and its theoretical analogue, dynamical maps, along with our concluding remarks.

\section{A brief Review}\label{secqpt}

\subsection{Quantum Process Tomography}

Quantum process tomography \cite{JModOpt.44.2455, PhysRevLett.78.390} is the experimental tool that determines the open evolution of a system that interacts with the surrounding environment. It is the tool that allows an experimenter to determine the unwanted action of a quantum process on the quantum bits going through it. It is an important tool for quantum information processing. A state going through a quantum gate or a quantum channel will experience some interactions with the surrounding environment.  Quantum process tomography allows the experimentalist to distinguish the differences between the ideal process and the process found experimentally.  Therefore it is an important tool in quantum control design and battling decoherence (loss of polarization).

The objective of quantum process tomography is to determine how a quantum process acts on different states of the system.  In very basic terms, a quantum process connects different quantum input states to different output states:
\begin{equation}
\mbox{input states} \rightarrow \mbox{process} \rightarrow \mbox{output states}.
\end{equation}
The complete behavior of the quantum process is known if the output state for any given input state can be predicted.

The tomography aspect of quantum process tomography is to use a finite number of input states, instead of all possible states, to determine the quantum process.  For instance, to determine a dynamical map, $\mathcal{B}$, we only need to know the mapping of the elements of the density matrix from an initial time to a final time,
\begin{eqnarray}
\mathcal{B}:\rho_{r's'}(t_0)\rightarrow\rho_{rs}(t).
\end{eqnarray}
The elements of the density matrix linearly span the whole state space \cite{modidis}.  

Experimentally, we do not have the access to the individual elements of the density matrix; we can only prepare physical states.  Thus, a set of physical states that linearly span the state space will be sufficient for the experiment.  
A state space of dimension $d$ requires $d^2$ states to span the space.  Once the evolution of each these input states is known, by linearity the evolution of any input state is known (see \cite{Nielsen00a} for detailed discussion).

Using the set linearly independent input states $P^{(m)}$, and measuring the corresponding output states $Q^{(m)}$, the evolution of an arbitrary input state can be determined. Let $\Lambda$ be the map describing the process, which we call \emph{process map}, and an arbitrary input state be expressed (uniquely) as a linear combination $\sum_j p^{(m)} P^{(m)}$.  The action of the map in terms of the matrix elements is as follows:
\begin{eqnarray*}
\sum_{r's'}\Lambda_{rr';ss'}\left(\sum_j p_j P_{r's'}^{(j)}\right) 
&=& \sum_j p_j Q_{rs}^{(j)}.
\end{eqnarray*}
With the knowledge of the output states corresponding to the input state we find the map by the following expression.
\begin{eqnarray}\label{sqptlin}
\Lambda_{rr';ss'} = \sum^{(m)} Q_{rs}^{(m)} {\tilde{P}}^{(m)^*}_{r's'},
\end{eqnarray}
where $\tilde{P}^{(n)}$ are the duals of the input states satisfying the scalar product 
\begin{eqnarray*}
{\tilde{P}}^{(m)^\dag}P^{(n)}=\sum_{rs}{\tilde{P}}^{(m)^*}_{rs}{P^{(n)}_{rs}} = \delta_{mn}.
\end{eqnarray*}

Today there are many variations of the quantum process tomography procedure described above, namely \emph{ancilla (entanglement) assisted process tomography} \cite{PhysRevLett.86.4195, PhysRevA.67.062307, PhysRevLett.90.193601, PhysRevLett.91.047902}, \emph{direct characterization of quantum dynamics} \cite{mohseni:170501,mohseni:062331}, \emph{selective efficient quantum process tomography}\cite{paz}, and \emph{symmetrized characterization of noisy quantum processes}\cite{laflamme}. Some of these procedures have been experimentally tested \cite{Nielsen:1998py, PhysRevA.64.012314, PhysRevLett.91.120402, Wein:121.13, orien:080502, NeeleyNature, chow:090502, Howard06, myrskog:013615}.  

\subsubsection{Weak coupling assumption}

In every quantum process tomography procedure listed above the input states are thought to be pure states.  There are two advantages of using pure states as inputs.  First, it is easier to span the space of the system with a set of pure state than it is with a set of mixed states. Second, pure states are always uncorrelated, which is one of the central assumption in every procedure above. This is often called the \emph{weak coupling assumption} \cite{PhysRevLett.75.3020}. It is simply a matter of preparing necessary pure states to perform a quantum process tomography experiment.

What if we depart from this assumption? It is well known that that initially correlated states can lead to not-completely positive dynamics for the system \cite{rodr,jordan:052110,jordan06a}.  In many recent experiments, the process maps that characterize the quantum operations have been plagued with negative eigenvalues and occasional non-linear behavior. We now examine the how the two cases are related.

The quantum process tomography procedures we reviewed require the input states be uncorrelated with the environment. Just before the experiment begins, in general the state of the system may be correlated with the environment. Thus, at the beginning of the experiment it is necessary to prepare the initially correlated total state into an uncorrelated state.  The difference between the process maps found from a quantum process tomography experiment and the dynamical maps calculated theoretically is precisely the act of preparation of input states.  Let us investigate this issue by analyzing the steps involved in a quantum process tomography experiment.

\subsection{Preparation procedure}\label{prep}

In practice, preparation procedures can be very complicated. Instead of describing many different procedures, we follow the theory of preparations developed in \cite{preppap}. Below is a brief review of the findings in that paper.

A preparation procedure is the mapping of a set of unknown states into a fix known input state. The most general transformation of a quantum state is described by a stochastic map \cite{SudarshanMatthewsRau61}. In light of that, a very complicated preparation procedure due to the aparatus, is simply denote by stochastic map, $\mathcal{A}^{(m)}$.  The procedure of preparing the $m$th input state is given as
\begin{eqnarray}\label{prepopen}
\rho^{\mathcal{SE}}\longrightarrow
\left[\mathcal{A}^{(m)}
\otimes\mathcal{I}\right]
\left(\rho^{\mathcal{SE}}\right)={R^\mathcal{SE}}^{(m)},
\end{eqnarray}
where $\mathcal{I}$ is the identity map acting on the state of the environment.  In other words, we are assuming that the preparation procedure acts only on the system\footnote{Generally we dot not require the preparation map to be trace preserving. However, the preparation procedure used here, the stochastic procedure, is trace preserving, hence we need to worry about this issue.  See \cite{preppap} for details.}. 

In general, the state of the system and the environment before preparation, $\rho^\mathcal{SE}$, will be correlated (see \cite{modigeo}).  The goal of the preparation procedure is to eliminate these correlations, as for an ideal experiment the state of the system should be uncorrelated with the environment at the beginning of the experiment. Therefore, under \emph{perfect preparation procedure assumption},  the right hand side of Eq. \ref{prepopen}, or the post-preparation procedure state, is of product form.

In \cite{preppap} two common preparation methods, that achieve product states. were discussed.  In this paper we will only focus on the method called \emph{stochastic preparation}.  The other method, \emph{projective preparations} will be discussed elsewhere.

\subsubsection{Stochastic preparation}\label{stoprep}

Many quantum experiments begin by initializing the system into a specific state. For instance, in the simplest case, the system can be prepared to the ground state by cooling it to near absolute zero temperature \cite{Wein:121.13, orien:080502, Howard06, myrskog:013615}. Mathematically, these set of operations are written as a pin map \cite{GoriniSudarshan},
\begin{eqnarray}
\Theta=\ket{\Phi}\bra{\Phi}\otimes\openone,
\end{eqnarray}
where $\openone$ is the identity matrix, acting as the `trace operator' ($\openone_{rs}\rho_{rs}=\mbox{Tr}[\rho]$) and $\ket{\Phi}$ is some fixed state (i.e. ground state) of the system. In this procedure, no matter what the initial state of the system was, it is ``pinned" to the final state $\ket{\Phi}\bra{\Phi}$.

The action of the pin map $\Theta$ on a bipartite state of the system and the environment.
\begin{eqnarray}\label{pinmap}
\Theta \left(\rho^\mathcal{SE}\right)
&=&\left[\Ket{\Phi}\Bra{\Phi}\otimes\openone \right]
\left(\rho^\mathcal{SE}\right)
=\Ket{\Phi}\Bra{\Phi}\otimes\rho^\mathcal{E}.
\end{eqnarray}
The pin map fixes the system into a single pure state, which means that the state of the environment is fixed into a single state as well; the pin map decorrelates the system from the environment. Once the pin map, $\Theta$, is applied, the system is prepared in the various different input states by applying local 
\begin{eqnarray}\label{stocmap}
\mathcal{A}^{(m)}\left(\rho^\mathcal{SE}\right)
&=&\left[\Omega^{(m)} \circ\Theta\right](\rho^\mathcal{SE})
= P^{(m)}\otimes\rho^\mathcal{E}.
\end{eqnarray}
As seen the last equations, the advantage of stochastic preparation method, as described here, is that the state of the environment is a constant for any input of the system.  This constancy of the state of the environment is  necessary to characterize the dynamical mechanisms properly \cite{modidis}. 

It may seem that stochastic preparation procedure alleviates the experiment from having an environment that depends on the preparation procedure. This is the main result of this paper, as we analyze this matter in much great detail below to show how inconsistencies in preparation procedure can lead to strange experimental results.

\subsection{Quantum process tomography experiment in steps}

The basic steps in a quantum process tomography experiment are broken down below:
\begin{itemize}
\item {Just before the experiment begins, the system and environment is in an unknown state $\rho^{\mathcal{SE}}$. The system and the environment in general are correlated; we are no longer making the weak coupling assumption.}
\item {The system is altered to a known input state by a preparation procedure.  The system and environment state after preparation is therefore given by 
$$\mathcal{A}^{(m)} \left(\rho^{\mathcal{SE}}\right)
\rightarrow P^{(m)}\otimes \rho^\mathcal{E} .$$ 
The input state is given by taking trace with respect to environment $$P^{(m)}=\mbox{Tr}_{\mathcal{E}}\left[P^{(m)}\otimes \rho^\mathcal{E} \right].$$}
\item {The system is then sent through a quantum process.  We consider the evolution to be a global unitary transformation in the space of the system \emph{and} the environment:
\begin{eqnarray*}
U P^{(m)}\otimes \rho^\mathcal{E} U^\dag.
\end{eqnarray*}}
\item {Finally the output state is observed.  Mathematically it is the trace with respect to the environment
\begin{eqnarray}\label{stocproceq}
Q^{(m)}=
\mbox{Tr}_{\mathcal E}\left[U P^{(m)}
\otimes \rho^\mathcal{E} U^\dag \right],
\end{eqnarray}
we will call this the 
\emph{stochastic process equation}.}
\item  {Finally using the input and the output states, we construct a map describing the process is constructed.}
\end{itemize}
The procedure above is identical to the procedure to find a dynamical map, except for the preparation procedure.  In the next two sections we will analyze quantum process tomography with the two preparation procedures discussed in last section.  The differences between what we find here and dynamical maps will be due to the preparation procedures (see Sec. \ref{seccon} for a discussion).  

\section{Quantum Process Tomography with Stochastic Preparations}\label{stochprep}

In this paper we analyze quantum process tomography procedures when the stochastic preparation procedure is used to generate the input states.  The stochastic preparation map's action on a bipartite state is given by Eq. \ref{stocmap}.  While the output states corresponding to the input is give by the stochastic process equation, Eq. \ref{stocproceq}.  All of our examples below will be based on these two equations.  Furthermore, all of our examples will have the same starting point, namely the unknown state before preparation
\begin{eqnarray}\label{totst}
\rho^{\mathcal{SE}}=\frac{1}{4}\left(\mathbb{I}\otimes\mathbb{I}+a_j\sigma_j\otimes\mathbb{I}+c_{23}\sigma_2\otimes\sigma_3\right).
\end{eqnarray}
This is a correlated state, but it is not an entangled state.  All examples will also have the same global dynamics in common, given by
\begin{eqnarray}\label{unitarys}
U = e^{-iHt} 
= \prod_{j}\left\{\cos\left(\omega t\right)\mathbb{I}\otimes\mathbb{I} -i\sin\left(\omega t\right)\sigma_j\otimes\sigma_j\right\},
\end{eqnarray}
where 
\begin{eqnarray}
H =  \omega\sum_{j=1}^{3} \sigma_j \otimes \sigma_j.
\end{eqnarray}
Let us now delve into constructing our examples. Starting with an ideal  second quantum process tomography experiment.

\subsection{Ideal stochastic preparation}\label{martinisT}

Let the pin map, $\Theta$ be
\begin{eqnarray}\label{pin2}
\Theta=\ket{\phi}\bra{\phi}\otimes\mathbb{I},
\end{eqnarray}
where $\ket{\phi}\bra{\phi}$ is a pure state of the system. The preparation of $\rho^{\mathcal{SE}}$ in Eq. \ref{totst} with this pin map leads to
\begin{eqnarray}
\Theta\left(\rho^{\mathcal{SE}}\right)=\ket{\phi}\bra{\phi}\otimes\frac{1}{2}\mathbb{I},
\end{eqnarray}
yielding the initial state $\ket{\phi}\bra{\phi}$ for the system qubit and a completely mixed state for the environment qubit. The next step is to create the rest of the input states using maps $\Omega^{(m)}$.  In this case, the fixed state $\ket{\phi}\bra{\phi}$ can be locally rotated to get the desired input state $P^{(m)}$ (given in Eq. \ref{prj})
\begin{eqnarray}\label{inp2}
\Omega^{(m)}\ket{\phi}\bra{\phi}\otimes\frac{1}{2}\mathbb{I}
&=& V^{(m)}\ket{\phi}\bra{\phi}{V^{(m)}}^{\dag}
\otimes\frac{1}{2}\mathbb{I}\nonumber\\
&=&P^{(m)}\otimes\frac{1}{2}\mathbb{I},
\end{eqnarray}
where $m=\{(1,-),(1,+),(2,+),(3,+)\}$ and $V^{(m)}$ are local unitary operators acting on the space of the system.  Where the inputs states are
\begin{eqnarray}\label{prj}
P^{(1,-)} = \frac{1}{2}\{\mathbb{I} - \sigma_1\},
&& P^{(1,+)} = \frac{1}{2}\{\mathbb{I} + \sigma_1\},\nonumber\\
P^{(2,+)} = \frac{1}{2}\{\mathbb{I} + \sigma_2\},
&& P^{(3,+)} = \frac{1}{2}\{\mathbb{I} + \sigma_3\}.
\end{eqnarray}

Now each input state is sent through the quantum process.  We will use the following unitary operator to develop the system and environment.
Noting the duals for the input states in Eq. \ref{prj} are
\begin{eqnarray}\label{dual}
&& \tilde{P}^{(1,-)} = \frac{1}{2}\{1 - \sigma_1- \sigma_2- \sigma_3\}, \nonumber\\
&& \tilde{P}^{(1,+)} = \frac{1}{2}\{1 + \sigma_1-\sigma_2-\sigma_3 \}, \nonumber\\
&& \tilde{P}^{(2,+)} =\sigma_2, \;\;\;
\tilde{P}^{(3,+)} =\sigma_3,
\end{eqnarray}
We can calculate the output states Eq. \ref{stocproceq}. The output states are
\begin{eqnarray}\label{stoout}
&& Q^{(1,-)}=\frac{1}{2}\{\mathbb{I}-\cos^2(2\omega t)\sigma_1\},\nonumber\\
&& Q^{(1,+)}=\frac{1}{2}\{\mathbb{I}+\cos^2(2\omega t)\sigma_1\},\nonumber\\
&& Q^{(2,+)}=\frac{1}{2}\{\mathbb{I}+\cos^2(2\omega t)\sigma_2\}, \nonumber\\
&& Q^{(3,+)}=\frac{1}{2}\{\mathbb{I}+\cos^2(2\omega t)\sigma_3\}.
\label{stochoutputs}
\end{eqnarray}

\begin{center}
\begin{figure}[t]
\resizebox{8.06 cm}{4.91 cm}{\includegraphics{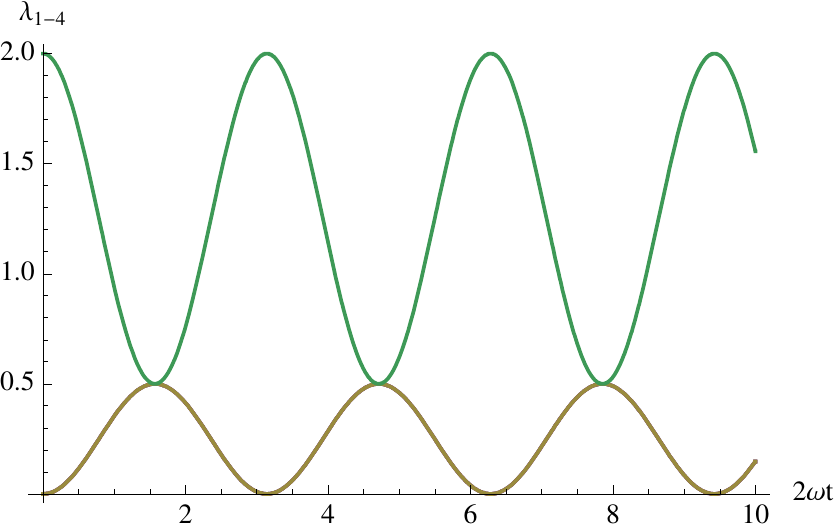}}
\caption{\label{fig5_1}
The eigenvalues of the process map in Eq. \ref{sact} are plotted as function of $2 \omega t$. As expected the eigenvalues are always positive. The stochastic preparation procedure allows the experimentalist to prepare any pure state for the system.  By convexity, then all possible states of the system are in the compatibility domain of the process map in Eq. \ref{sact}.}
\end{figure}
\end{center}

The linear process map is constructed using Eq. \ref{sqptlin}, the duals in Eq. \ref{dual}, and the output states:
\begin{eqnarray}\label{sact}
\Lambda_s=\frac{1}{2}
\left(\begin{array}{cccc}
1+C^2 &0&0&2C^2\\
0& 1-C^2&0 & 0\\
0 & 0& 1-C^2 &0\\
2C^2& 0&0& 1+C^2
\end{array}\right),
\end{eqnarray}
where $C=\cos(2\omega t)$.  The eigenvalues of the process map are plotted in Fig. \ref{fig5_1}. The map above is completely positive and can be tested for linearity.  This would be the result of an ideal quantum process tomography experiment (see Sec. \ref{martinis} for such an experiment).  The key thing note above is that the state of the environment is completely constant throughout the problem.  Below we will show that this need not be the case all of the times.

\subsection{Negative maps due to control errors}\label{chowT}
Suppose the initial state is prepared well using a pin map
\begin{eqnarray}\label{pspure2}
\Theta\left(\rho^{\mathcal{SE}}\right)&=&P^{(3,+)}\otimes\mathbb{I}.
\end{eqnarray}
After obtaining this state, the other input states are prepared by local rotations.  Let us consider the case where one of the rotation is not perfect.
\begin{eqnarray}
V^{(1,-)}\ket{1}
\rightarrow\frac{1}{\sqrt{2}} \left(
\sqrt{1-\epsilon}\ket{1}-\sqrt{1+\epsilon}\ket{0}\right),
\end{eqnarray}
where $\epsilon$ is taken to be a small positive real number.  We introduced a small error for the preparation of $P^{(1,-)}$, but we have kept the error simple by not giving it an additional phase, i.e. keeping $\epsilon$ to be real.  For simplicity we have also assumed that the error of this sort occurs in the preparation of only one state.

Let us now pretend that we are not aware of this error.  Then in reality we have the following set of input states
\begin{eqnarray}\label{psprj2}
P^{(1,-)}&=&\frac{1}{2}\{\mathbb{I}
+\epsilon\sigma_3
-\sqrt{1-\epsilon^2}\sigma_1\}, \nonumber\\
P^{(1,+)}&=&\frac{1}{2}\{\mathbb{I}+\sigma_1\},\nonumber\\
P^{(2,+)}&=&\frac{1}{2}\{\mathbb{I}+\sigma_2\}, \nonumber\\
P^{(3,+)}&=&\frac{1}{2}\{\mathbb{I}+\sigma_3\}.
\end{eqnarray}
Let us use the same unitary evolution as before given in Eq. \ref{unitarys}. The output states corresponding to the last three input states in Eq. \ref{psprj2} are the same as before, given by Eq. \ref{stochoutputs}.  For the input state $P^{(1,-)}$, the corresponding output state is follows
\begin{eqnarray}\label{poorout}
Q^{(1,-)}&=&\frac{1}{2}\{\mathbb{I}
+(\epsilon\sigma_3
-\sqrt{1-\epsilon^2}\sigma_1)\cos^2(2\omega t)\},\nonumber\\
Q^{(1,+)}&=&\frac{1}{2}\{\mathbb{I}+\cos^2(2\omega t)\sigma_1\},\nonumber\\
Q^{(2,+)}&=&\frac{1}{2}\{\mathbb{I}+\cos^2(2\omega t)\sigma_2\},\nonumber\\
Q^{(3,+)}&=&\frac{1}{2}\{\mathbb{I}+\cos^2(2\omega t)\sigma_3\}.
\end{eqnarray}
Using these output states, the duals given by Eq. \ref{dual}, and Eq. \ref{sqptlin}, we can find the process map.  

\begin{center}
\begin{figure}[!ht]
\resizebox{8.06 cm}{4.91 cm}
{\includegraphics{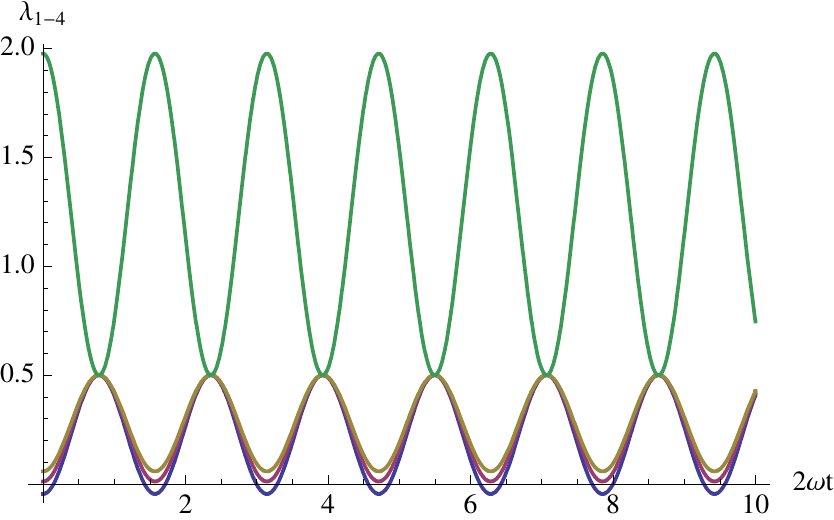}}
\caption{\label{negfig}
The eigenvalues of the process map found using the output states in Eq. \ref{poorout} and the duals in Eq. \ref{dual}.  One of the eigenvalue is negative for certain times; we have taken $\epsilon=0.1$.  The negativity is due to the errors in the unitary operation implemented to prepare one of the input states.}
\end{figure}
\end{center}

The process map turns out to be rather complicated, and its eigenvalues are even more complicated looking.  Therefore, we do not write them down, instead we have plotted the eigenvalues as function of $2\omega t$.  We take the value for the error to be $\epsilon=0.1$ for the plot in Fig \ref{negfig}.  One of the eigenvalue in Fig. \ref{negfig} is negative for certain times.  This shows yet another cause for negative eigenvalues in a process map. The negative eigenvalues here have nothing to do with the initial correlations between the system and the environment.  The negative eigenvalues are attributed to poor control in the preparation procedure.

Now we have discussed several scenarios that can lead to negative eigenvalues for a process map.  Two comments are in order at this point.  When the prepared input states are not pure and the process map has negative eigenvalues, then one should be weary of initial correlations with the environment.  To further check this, the process map should be tested for linearity by sending several additional input states through the process.  If the process map predicts the output states properly, then one can be confident that the system is initially correlated with the environment.  If the process map does not predict the output states correctly (non-linear behavior), then there are additional problems with the experiment, including the possibility of poor control in the preparation procedure.  In the case where the input states are pure and the process map has negative eigenvalues, the negativity can only come from either inconsistencies in the preparation procedure or poor preparation control.

\subsection{Mixed input states}\label{howardT}

Demanding pure input states in a quantum process tomography experiment guarantees initially uncorrelated state of the system.  Though if it is not possible to prepare pure states, we can still determine the process. Below we discuss what happens when the input is not a pure state.  We will show that this creates an uncertainty about whether the system is correlated with the environment or not.  Yet we show that regardless of this uncertainty, the quantum process tomography experiment can be carried out consistently.

\subsubsection{Uncorrelated}
Suppose the initial state is prepared by a pin map leading to
\begin{eqnarray}\label{pspure1}
\Theta\left(\rho^{\mathcal{SE}}\right)&=&
\left(pP^{(3,+)}+(1-p)P^{(3,-)}\right)\otimes\mathbb{I}\\
&=&\frac{1}{2}\{\mathbb{I}+p\sigma_3\}\otimes\mathbb{I},
\end{eqnarray}
where $0<<p<1$.  The rest of the input states are prepared by rotations.  Then the input states are
\begin{eqnarray}\label{psprj}
P^{(1,-)}=\frac{1}{2}\{\mathbb{I}-p\sigma_1\},&&
P^{(1,+)}=\frac{1}{2}\{\mathbb{I}+p\sigma_1\},\\
P^{(2,+)}=\frac{1}{2}\{\mathbb{I}+p\sigma_2\},&&
P^{(3,+)}=\frac{1}{2}\{\mathbb{I}+p\sigma_3\}.
\end{eqnarray}
For the unitary operator in Eq. \ref{unitarys}, the corresponding output states will be
\begin{eqnarray}
&& Q^{(1,-)}=\frac{1}{2}\{\mathbb{I}-p\cos^2(2\omega t)\sigma_1\},\nonumber\\
&& Q^{(1,+)}=\frac{1}{2}\{\mathbb{I}+p\cos^2(2\omega t)\sigma_1\},\nonumber\\
&& Q^{(2,+)}=\frac{1}{2}\{\mathbb{I}+p\cos^2(2\omega t)\sigma_2\},\nonumber\\
&& Q^{(3,+)}=\frac{1}{2}\{\mathbb{I}+p\cos^2(2\omega t)\sigma_3\}.
\end{eqnarray}

The only change that we have to make to find the process map is to define a dual proper set, in this case
\begin{eqnarray}\label{psdual}
&& \tilde{P}^{(1,-)} = \frac{1}{2p}\{p\mathbb{I} -\sigma_1-\sigma_2- \sigma_3\}, \nonumber\\
&& \tilde{P}^{(1,+)} = \frac{1}{2p}\{p\mathbb{I} +\sigma_1-\sigma_2-\sigma_3 \}, \nonumber\\
&&\tilde{P}^{(2,+)} =\frac{1}{p}\sigma_2,\;\;\;
\tilde{P}^{(3,+)} =\frac{1}{p}\sigma_3. 
\end{eqnarray}
We can find the process map using Eq. \ref{sqptlin}.  
\begin{eqnarray}
\Lambda_{mx1}=\frac{1}{2}
\left(\begin{array}{cccc}
1+C^2 &0&0&2C^2\\
0& 1-C^2&0 & 0 \\
0 & 0& 1-C^2 &0 \\
2C^2& 0&0& 1+C^2
\end{array}\right),
\end{eqnarray}
where $C=\cos(2\omega t)$.  The process map here case turns out to be the same as in Eq. \ref{sact}.  This is to expected; the process map for stochastic preparation is defined over set of all states, including the input states above.

\subsubsection{Correlated}
The downside of course is that there is no way to distinguish the states in Eq. \ref{pspure1} from the following state
\begin{eqnarray}
\Theta\left(\rho^{\mathcal{SE}}\right)=\frac{1}{2}\{\mathbb{I}\otimes\mathbb{I}+p\sigma_3+c_{23}\sigma_2\otimes\sigma_3\}.
\end{eqnarray}
Unlike in Eq. \ref{psprj}, this is a correlated state. Yet, both total states have the same reduced state for the system part.

Even in this case we can find the process map properly if we use the correct dual set given in Eq. \ref{psdual}.  The process map in that case will be 
\begin{eqnarray}\label{dynamicalmap}
\Lambda_{mx2}=\frac{1}{2}\left(
\begin{array}{cccc}
1+C^2 & 0 & -c_{23} CS & 2 C^2 \cr
0 & 1-C^2 & 0 &-c_{23} CS \cr
-c_{23} CS & 0 & 1-C^2  & 0 \cr
2C^2& -c_{23} CS & 0 &1+C^2 \cr 
\end{array}\nonumber
\right),
\end{eqnarray}
where $C=\cos(2\omega t)$ and $S=\sin(2\omega t)$. Which the same as the dynamical map calculated in (Eq. 6) \cite{CesarEtal07}. In both of the examples above, had we assumed that the input states were close enough to the pure states we desired, and used the dual set give by Eq. \ref{dual}, then the process map would contain an error.

\begin{center}
\begin{figure}[!ht]
\resizebox{8.06 cm}{4.91 cm}{\includegraphics{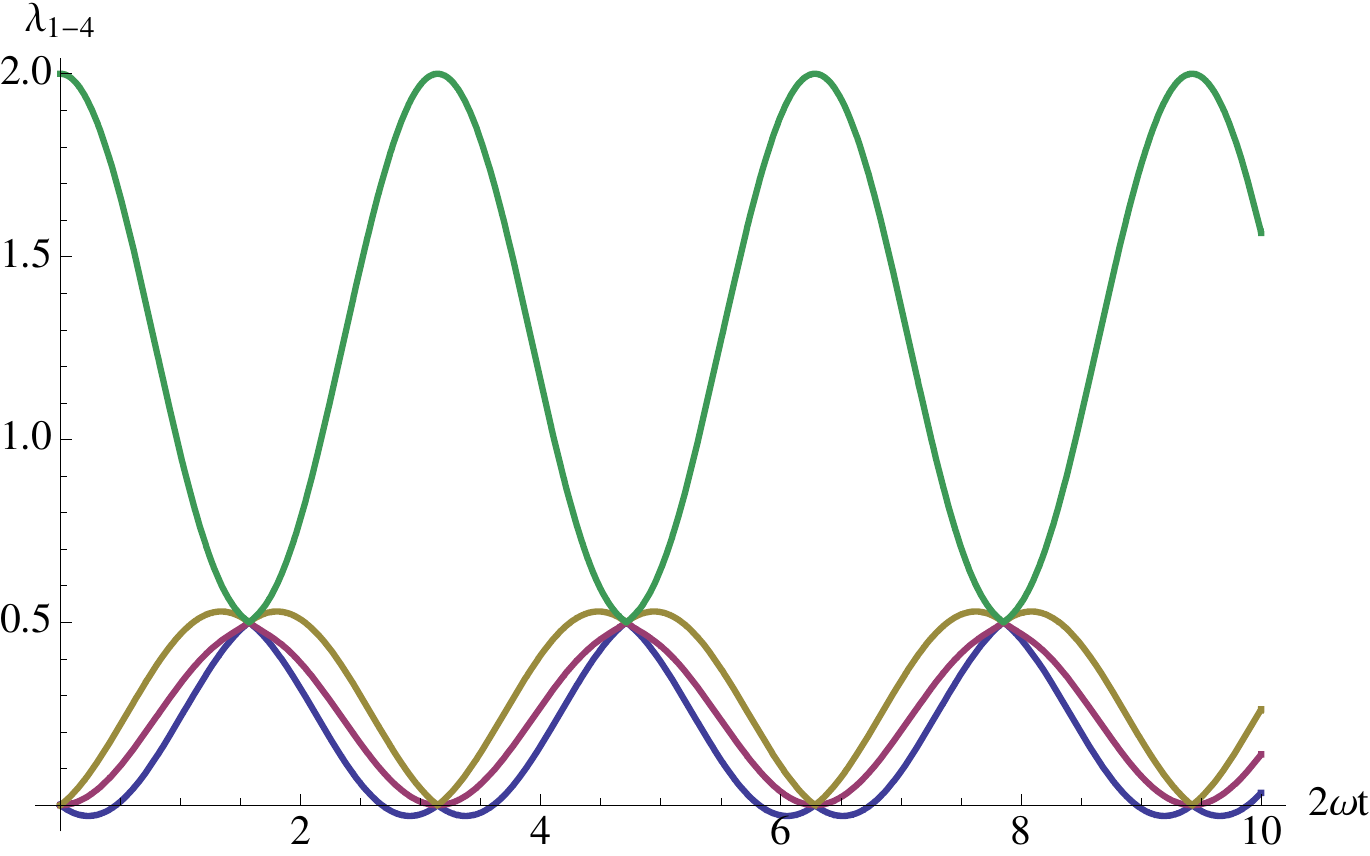}}
\caption{\label{fig2_1}
The eigenvalues of the dynamical map in Eq \ref{dynamicalmap} are plotted as function of $2 \omega t$.  One of the eigenvalue is negative for certain values of $\omega t$; we have taken $c_{23}=0.5$.  The negative eigenvalue is due to the initial correlations between the system and the environment.}
\end{figure}
\end{center}

These two examples illustrate the contrary argument to what some have suggested, that only the completely positive process maps should be considered physically valid.  We showed in the second example that, one can obtain a not-completely positive process map in a consistent fashion.  The real issue is that the correlations with environment are not convenient for experimental purposes.  But that does not mean that one should fix a not completely positive process map to a completely positive process map with numerical methods \cite{havel03,ziman}.

\subsection{Multiple stochastic preparations} \label{myrskogT}

The pin map used in stochastic preparation must be used consistently.  Let us show an example of what happens when two stochastic preparation procedures are used.  Consider a quantum process tomography experiment where the following linearly independent states are used to span the space of the system
\begin{eqnarray}
P^{\mathbb{I}}=\frac{1}{2}
\mathbb{I},\;\;P^{(1+)},\;\;P^{(2+)},\;\;P^{(3+)}.
\end{eqnarray}
These states form the linearly independent set\footnote{The linear combination will not always be convex.  For example $P^{(2,-)}=P^{(1,+)}+P^{(1,-)}-P^{(2,+)}$.  Also notice that these four states form a linearly independent set, but they are not orthogonal to each other.} $\{\mathbb{I},\sigma_j\}$; which is different from the set used in previous examples

Consider the following two qubit state as the available state to the experimenter at $t=0_-$:  This is the same state we used in Eq. \ref{totst} for the previous example.

Let the pin map, $\Theta$, for our example be
\begin{eqnarray}\label{pin1}
\Theta=\ket{\phi}\bra{\phi}\otimes\mathbb{I},
\end{eqnarray}
where $\ket{\phi}\bra{\phi}$ is a pure state of the system. The preparation of $\rho^{\mathcal{SE}}$ with this pin map leads to
\begin{eqnarray}\label{inp1}
\Theta\left(\rho^{\mathcal{SE}}\right)=\ket{\phi}\bra{\phi}\otimes\frac{1}{2}\mathbb{I},
\end{eqnarray}
yielding the initial state $\ket{\phi}\bra{\phi}$ for the system qubit and a completely mixed state for the environment qubit. The next step is to create the rest of the input states using maps $\Omega^{(m)}$.  In this case, the fixed state $\ket{\phi}\bra{\phi}$ can be locally rotated to get the desired input state $P^{(m)}$ (given in Eq. \ref{prj})
\begin{eqnarray}\label{inp3}
\Omega^{(m)}\ket{\phi}\bra{\phi}\otimes\frac{1}{2}\mathbb{I}
&=& V^{(m)}\ket{\phi}\bra{\phi}{V^{(m)}}^{\dag}
\otimes\frac{1}{2}\mathbb{I}\nonumber\\
&=&P^{(m)}\otimes\frac{1}{2}\mathbb{I},
\end{eqnarray}
where $m=\{(1,-),(1,+),(2,+),(3,+)\}$ and $V^{(m)}$ are local unitary operators acting on the space of the system.

Once again, we take the same initial state and unitary operators as in the last example.  Now suppose the pin map given in Eq. \ref{pin1} is used to prepare the state $\ket{\phi}\bra{\phi}$, and then by local transformations  $P^{(1,+)}$, $P^{(2,,+)}$, and $P^{(3,+)}$ are prepared. Finally the mixed state is prepared by letting $P^{(3+)}$ decohere.  This is a different pin map than the one in Eq. \ref{pin1}, so we are using two pin maps to prepare two sets of input states.  The unitary operator, we are using, is often called the \emph{swap gate}, because it swaps the states of two qubits with the period of $t=\frac{\pi}{4\omega}$.  Then at $t=\frac{\pi}{4\omega}$, the total state will be
\begin{eqnarray}
\rho^{\mathcal{SE}}\left(\frac{\pi}{4\omega}\right)
=\frac{1}{4}
\{\mathbb{I}\otimes\mathbb{I}
+a_j \mathbb{I}\otimes\sigma_j\}.
\end{eqnarray}
The state of the system has fully decohered.  

The corresponding output states for the input states above are found using Eq. \ref{stocproceq}. The state of environment in that equation for input $P^{\mathbb{I}}=\frac{1}{2}\mathbb{I}$ is $\rho^{\mathcal E}=\frac{1}{2} \{\mathbb{I} +a_j\sigma_j\}$, while for the other inputs the state of the environment $\rho^{\mathcal E}=\frac{1}{2}\mathbb{I}$.  Then the corresponding output states are
\begin{eqnarray}\label{multstocout}
&& Q^{(\mathbb{I})}=\frac{1}{2}\{\mathbb{I}+\sin^2(2\omega t)\sigma_3\}, \nonumber\\
&& Q^{(1,+)}=\frac{1}{2}\{\mathbb{I}+\cos^2(2\omega t)\sigma_1\},\nonumber\\
&& Q^{(2,+)}=\frac{1}{2}\{\mathbb{I}+\cos^2(2\omega t)\sigma_2\}, \nonumber\\
&& Q^{(3,+)}=\frac{1}{2}\{\mathbb{I}+\cos^2(2\omega t)\sigma_3\}.
\end{eqnarray}
The last three output states are the same as in the last example, but the fourth one is different.

Now suppose we calculate the output state corresponding to the input $P^{(-1)}$. By linearity we have $P^{(-1)}=\mathbb{I}-P^{(+1)}$.  If the process is linear, then the output state for this input state is given by
\begin{eqnarray}
Q^{(-1)}&=&2 Q^{(\mathbb{I})}-Q^{(+1)}\nonumber\\
&=&\frac{1}{2}\{\mathbb{I}+2\sin^2(2\omega t)\sigma_3
-\cos^2(2\omega t)\sigma_1\}.
\end{eqnarray}
This state is not physical for certain times, and therefore the linearity and the positivity of the process is violated.  

\begin{center}
\begin{figure}
\resizebox{8.06 cm}{4.91 cm}
{\includegraphics{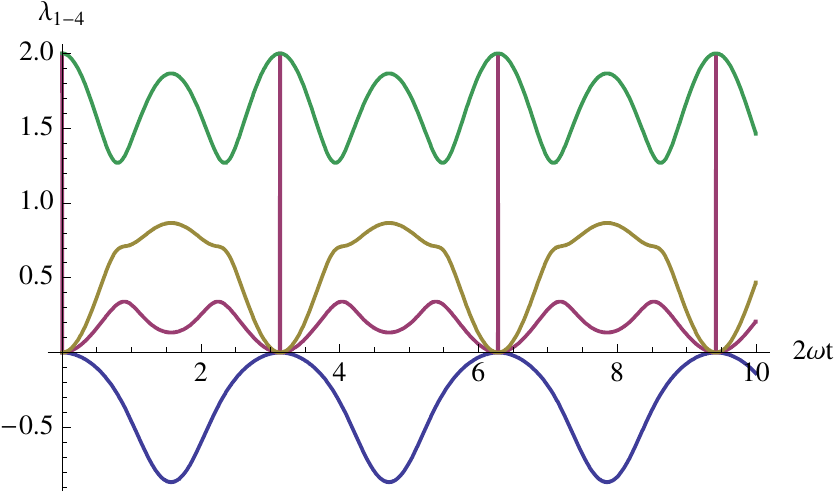}}
\caption{\label{fig5_2}
The eigenvalues of the process map in Eq. \ref{msact} plotted as function of $2 \omega t$. One of the eigenvalue is negative for certain values of $\omega t$.  The negativity is due to the inconsistency in the preparation procedure, and not due to the initial correlations with the environment.}
\end{figure}
\end{center}

The process has not changed from the last example, only the method of determining the process has.  Thus when the stochastic map is not used consistently, the process map can behave nonlinear.  For completeness we find the process map for this example,
\begin{widetext}
\begin{eqnarray}\label{msact}
\Lambda_{ms}=\frac{1}{2}
\left(\begin{array}{cccc}
1+C^2 &-(1+i)S^2&0&2C^2\\
-(1-i)S^2& 1-C^2+2S^2&0 & 0\\
0 & 0& 1-C^2 &(1+i)S^2\\
2C^2& 0&(1-i)S^2& 1+C^2-2S^2
\end{array}\right),
\end{eqnarray}
\end{widetext}
where $C=\cos(2\omega t)$ and $S=\sin(2\omega t)$. The eigenvalues of this process map are negative, as seen in Fig. \ref{fig5_2}. The fact that the state of the environment is not a constant for all of the input states leads to the negativity and non-linearity of the process map.

The origin for the negative eigenvalues here is completely different than the negative eigenvalues is not due to the initial correlations between the system and the environment. Since none of the inputs were correlated with the environment here, negativity arises from the inconsistencies in the preparation procedure.

While this example may not seem realistic, the point regarding inconsistency arising from multiple pin maps still stands.  In realistic cases, the trouble may not be seen so easily, due to complicated interactions with the environment. We will discuss an experiment where multiple stochastic preparation procedures are implemented in Sec. \ref{myrskog}.

\section{Analysis of experiments}\label{secexp}

In this section we analyze four quantum process tomography experiments that are analogues of the four constructed examples of the last section. Each experiment is a well performed experiment in its own right.  We analyze each experimental procedure and comment on why the result was not completely positive (when applicable).

\subsection{Quantum process tomography of Josephson-phase qubit and two-level state}\label{martinis}

In this experiment \cite{NeeleyNature}, a Josephson-phase qubit along with a more stable but less controllable two-level state acting as memory are examined.  The Josephson-phase is initially prepared into the ground state with high fidelity.  Which is then prepared into various excited states via unitary transformations.  Finally it is `swapped' with the memory qubit and then some time later swapped back.  Quantum process tomography of this procedure is implemented. By doing so, they observe that the decoherence of the Josephson-phase qubit can be  hindered by transferring the information to the stable two-level state.

The experimentalist find the resulting process map to be completely positive \emph{without} any numerical corrections.  This in principle is an ideal quantum process tomography experiment.  The absence of negativity from the process map suggests that the system of interest was not in anyway correlated with the surrounding degrees of freedom.  

However to be sure, one should check this process map for linearity by preparing arbitrary initial states and putting them through the swap procedure. If the experimental result matches the theoretical predictions of the process map, then the results of this experiment fit the analysis of the constructed example in Sec. \ref{martinisT}.  

\subsection{Quantum process tomography solid-state qubit}\label{chow}

In this experiment \cite{chow:090502}, a solid-state qubit is put through various gates.  Quantum process tomography of these gates is performed.  They perform quantum process tomography on three operations; identity, rotations along $x-$plane, and rotations along $y-$plane. The results are excellent, with very little errors. 

For the case where identity operation is made, the process map is nearly identity as expected.  This means that the input states are well prepared, and are nearly pure. For the other transformations, the errors are attributed to errors in preparation procedures and subsequent measurements. 

The process map obtained in each case, however has negative eigenvalues.  Since the negativity is along the same magnitude for the case when no operation is made and when rotations along $x$ and $y-$planes are implemented, we can presume that the errors must be due to the errors in unitary operations used to prepare input states.  This is along the same lines as the example in Sec. \ref{chowT}.

\subsection{Quantum process tomography in NMR}\label{howard}

In this experiment \cite{Howard06}, the system that is studied is an electron configuration formed in a nitrogen vacancy defect in a diamond lattice.  The quantum state of the system is given by a spin triplet ($S=1$).  Again we will write the initial state of the system and environment as $\rho^\mathcal{SE}$.  

The system is prepared by optical pumping, which results in a strong spin polarization.  The state of the system is said to have a 70\% chance of being in a pure state $\ket{\phi}$.  Or more mathematically, the probability of obtaining $\ket{\phi}$ is $\mbox{Tr}[\ket{\phi}\bra{\phi} \rho^\mathcal{SE}] = 0.7$.  

Since the population probability is high, an assumption was made that the state of the system can be simply approximated as a pure state $\ket{\phi}\bra{\phi}$.  From this initial state, different input states can be prepared by suitably applying microwave pulses resonant with the transition levels.  After preparation, the system is allowed to evolve, and the output states are determined by quantum state tomography.  With the knowledge of the input state and the measured output states, a linear process map is constructed.

It was found that the linear process map has negative eigenvalues, so the map was ``corrected" using a least squares fit between the experimentally determined map and a theoretical map based on Hermitian parameterization \cite{havel03}, while enforcing complete positivity. 

However, if we do not regard the negative eigenvalues of the map as aberrations, then we should consider the assumptions about the preparation of the system more carefully.  The assumption about the initial state of the system is:
\begin{eqnarray}
\rho^\mathcal{SE} \rightarrow \ket{\phi}\bra{\phi}\otimes\tau .
\end{eqnarray}
This is in effect a  pin map.  Along with the pin map, the stochastic transformations are applied on the initial state to prepare the various 
input states; this is identical to the stochastic preparation method discussed in Sec. \ref{stoprep}.

It is clear that the pure initial state assumption is unreasonable given our knowledge now of how the process is sensitive to the initial correlations between the system and the environment.  In effect the action of the pin map in this experiment is not perfect, and the pin map can be ignored.  Then the process equation is:
\begin{eqnarray}
Q^{(m)} = \mbox{Tr}_\mathbb{B} [U \Omega^{(m)} \rho^\mathcal{SE} U^\dagger] 
\end{eqnarray}
where $\Omega^{(m)}$ is the stochastic transformation that prepares the $m$th input state. In this experiment, $\Omega^{(m)}$ is nothing more than a unitary transformation $V^{(m)}$ satisfying $V^{(m)} \ket{\phi} = \ket{\psi^{(m)}}$, where $\ket{\psi^{(m)}}$ is the desired pure $m$th input state.

We can write the unitary transformation for a two-level system as:
\begin{eqnarray}
V^{(m)} = \ket{\psi^{(m)}}\bra{\phi}+ \ket{\psi^{(m)}_\perp }\bra{\phi_\perp}
\end{eqnarray}
where $\braket{\psi^{(m)}|\psi^{(m)}_\perp } = \braket{\phi | \phi_\perp } = 0$.  This defines $V^{(n)}$ as a transformation from the basis $\{\ket{\phi}\}$ to the basis $\{\ket{\psi^{(n)}_i}\}$.  The equation for the process becomes:
\begin{eqnarray*}
Q^{(m)} &=& 
\mbox{Tr}_\mathcal{E} \left[U \ket{\psi^{(m)}}\braket{\phi| 
\rho^\mathcal{SE} |\phi}\bra{\psi^{(m)}} U^\dagger\right]\\
&& +  \mbox{Tr}_\mathcal{E} \left[U \ket{\psi^{(m)}_\perp}\braket{\phi_\perp| \rho^\mathcal{SE} |\phi}\bra{\psi^{(m)}} U^\dagger\right]\\
&& +  \mbox{Tr}_\mathcal{E} \left[U \ket{\psi^{(m)}}\braket{\phi| \rho^\mathcal{SE} |\phi_\perp}\bra{\psi^{(m)}_\perp} U^\dagger\right]\\
&& +  \mbox{Tr}_\mathcal{E} \left[U \ket{\psi^{(m)}_\perp}\braket{\phi_\perp| \rho^\mathcal{SE} |\phi_\perp}\bra{\psi^{(m)}_\perp} U^\dagger\right].
\end{eqnarray*}
Therefore, since $\braket{\phi | \rho^\mathcal{SE} | \phi} = 0.7$ to first approximation, the process is a linear mapping on the states $\ket{\psi^{(m)}}\bra{\psi^{(m)}}$.  However, it is clear that if all terms are included, the process is not truly linear in the states $\ket{\psi^{(m)}}\bra{\psi^{(m)}}$.  The  negative eigenvalues are therefore a result of fitting results into a linear map when the process is not truly represented by a linear map.

There is no way to tell if the system in the experiment above was initially correlated with the environment or not, as shown in Sec. \ref{howardT}.  However the negativity in the process map obtained that the system may have been correlated initially.  If this was the case the obtained map may be linear and well behaved.  Fixing it numerically to make it completely positive may in reality make it behave in non linear fashion.

\subsection{Quantum process tomography of motional states}\label{myrskog}

In this experiment \cite{myrskog:013615}, quantum process tomography of the motional states of  trapped $^{85}Rb$ atoms in the potential wells of a one dimensional optical lattice is performed.  Only two bound bands are considered, which are labeled as states $\ket{0}$ and $\ket{1}$. The states are prepared stochastically.  An initial state of the system is the ground state $\ket{g}\bra{g}$, and from it states $\ket{r}\bra{r}$, $\ket{i}\bra{i}$, and the fully mixed state $\frac{1}{2}\mathbb{I}$ are prepared.  The states $\ket{r}$ and $\ket{i}$ stand for the real and imaginary coherence states, which are prepared by applying appropriate unitary transformations on the ground state.  This is achieved by displacing the lattice for the real coherence state and for the imaginary coherence state, a quarter-period delay is added after the displacement. The identity state is prepared by letting a superposition state decohere. The input states are allowed to evolve and the output states are determine by quantum state tomography.

The process map is found following the usual procedure laid out in Sec. \ref{secqpt}.  Since there are particles lost to the neighboring cells, the map is not required to be trace preserving.  Based on this loss they also argue that the map can pick non-physical behavior (not-completely positive).  The map is forced to be ``physical" (completely positive) by using the maximum likelihood method \cite{Ziman06}.

In our terminology, the states prepared are $P^{(1,+)}$, $P^{(2,+)}$, $P^{(3,-)}$, and $\frac{1}{2}\mathbb{I}$.  The three projective states are prepared by a single consistent stochastic preparation, while the fully mixed state is prepared by letting the state $P^{(1,+)}$ decohere. Which means an additional pin map is applied to prepare one of the input states.  
As we saw in the example in Sec. \ref{myrskogT}, this can lead to a not-completely positive and non-linear process map.

Furthermore, the input states have varying values for polarization. Their data is listed in the table below.
\begin{center}
\begin{tabular}{l | llll}
\hline
& $\rho_g$ & $\rho_{\mathbb I}$ & $\rho_r$ & $\rho_i$ 
\\ \hline 
$P^{(3,-)}$ & 0.90 & 0.60 & 0.69 & 0.69 \\
$P^{(3,+)}$ & 0.10 & 0.40 & 0.31 & 0.31 \\
$P^{(1,+)}$ & 0.82 & 0.59 & 0.85 & 0.63 \\
$P^{(2,+)}$ & 0.84 & 0.58 & 0.64 & 0.37 \\
\hline
\end{tabular}
\end{center}
where $\rho_j$ are the experimentally prepared  states projected onto projectors $P^{(m)}$. The polarization of the imaginary state in the $\sigma_2$ direction is very low.  This could mean that it is correlated with the environment, while the polarization of the ground state along the negative $\sigma_3$ direction is almost unity, meaning it is only weakly correlated at best.

Here, we have two potential causes for the process map to have negative eigenvalues.  The first problem is with the experimental procedure; applying multiple stochastic preparations that may affect the state of the environment differently.  The second problem may be unavoidable; since it is extremely difficult to prepare pure states in a setup like this.  Even if the negative eigenvalues are due to the initial correlations, we do not have a prescription to obtain a process map in a consistent fashion.  For the last example shown in Sec. \ref{stochprep}, we assumed that the initial correlations were constant throughout the process.  This does not seem to be the case here. Since the ground state is almost pure, while the imaginary state is clearly not pure.  Therefore the correlation with environment for these two inputs must be different as in the case of the example in Sec. \ref{myrskogT}.

\section{Conclusion}\label{seccon} 

The examples with multiple stochastic preparations (in Sec. \ref{myrskogT}) and projective preparations (in Sec. V.A) in \cite{kuah:042113}) look very similar.  For multiple stochastic preparations, the state of the environment depends on the stochastic map.  If we think of each projective preparations as an independent stochastic preparation then the state of the environment, in the example (in Sec. V.A) in \cite{kuah:042113}, depends on the preparation map.  Then are the two situations the same? Let us show below that the two situations are fundamentally different. 

In the weak coupling limit, the projective preparation procedure does not play an important role.  As we saw in the last section, when the system and the environment are initially uncorrelated, the projective preparation procedure will have no affect on the state of the environment. This is because the state of the environment is affected only indirectly due to the initial correlations between the system and the environment.  Thus the projective preparations, in the weak coupling limit, yield a linear process map.

This is not the case in the multiple stochastic preparation example. When multiple stochastic maps are used to prepare different input states, the inconsistencies do not stem from the initial correlations between the system and the environment.  In fact, in our example all input states are initially uncorrelated from the environment.  The inconsistencies arise from the preparation procedures themselves, leading to different states of the environment for different input states.  These inconsistencies will be absent in the case where the system develops in a closed form, since the system does not feel the presence of the environment during the quantum process.  Hence in the ``weak interaction limit'' multiple stochastic preparations yield a linear process map.

Lastly, note that each of the process map found in this section, given by Eqs.  \ref{msact}, and (in Sec. V.A) in \cite{kuah:042113} are different from each other.  This clearly shows that the preparation procedures play a non-trivial role in open quantum system experiments.  The preparation procedure is the only thing that distinguished each case.  For the case where no preparation procedure is applied, i.e. the case of the dynamical map (Eq. 6 in \cite{CesarEtal07}), the situation is still different.  The dynamical map, which has negative eigenvalues, is linear and has a valid interpretation within the compatibility domain.  The process maps in Eqs. \ref{msact} and (in Sec. V.A) in \cite{kuah:042113} do not have any consistent interpretation. 

When negative eigenvalues are found in a process map, they hint to some problem in the preparation procedure.  Although this is a bit premature to state at this point; we analyze the negative eigenvalues in process map in more detail in the next section.

Our study of quantum process tomography started by noting that the dynamical map acting on a system can have negative eigenvalues.  The dynamical map has negative eigenvalues when the system is initially correlated with the environment.  In the course of our studies of quantum process tomography, we showed that the preparation procedure cannot be  neglected for any quantum system that interacts with an environment.  These are the two major themes discussed in this paper. Though, along the way,  we presented a method of quantum process tomography that is independent of the preparation procedure.  The map arising from this procedure lead us to an expression that quantifies the memory effect on the dynamics of the system due to the initial  correlations with the environment. Determining the memory effect is an important task in coherence control.

{\bf Acknowledgments:} I am grateful to Aik-Meng Kuah, Ali Rezakhani, C\'esar Rodr\'iguez-Rosario, and Daniel Terno for valuable conversations. This work was financially supported by the National Research Foundation and the Ministry of Education of Singapore.

\bibliography{dissertation} 										 
\end{document}